\documentclass[aps,prb,twocolumn,groupedaddress,showpacs,superscriptaddress]{revtex4}
\usepackage{graphicx}
\usepackage{epsfig}

\begin{document}

\title{Negative refractive index due to chirality}

\author{Jiangfeng Zhou}  
\affiliation{Ames Laboratory and Department of Physics and
Astronomy, Iowa State University, Ames, Iowa 50011, USA}
\author{Jianfeng Dong}
\affiliation{Ames
Laboratory and Department of Physics and Astronomy, Iowa State
University, Ames, Iowa 50011, USA}
\affiliation{Institute of Optical Fiber Commun. and Network Tech.,
Ningbo University, Ningbo 315211, China}
\author{Bingnan Wang}
\affiliation{Ames Laboratory and Department of Physics and
Astronomy, Iowa State University, Ames, Iowa 50011, USA}

\author{Thomas Koschny}

\affiliation{Ames Laboratory and Department of Physics and
Astronomy, Iowa State University, Ames, Iowa 50011, USA}
\affiliation{Institute of Electronic Structure and Laser -
Foundation for Research and Technology Hellas (FORTH),and Department
of Materials Science and Technology, University of Crete, Greece}

\author{Maria Kafesaki}
\affiliation{Institute of Electronic Structure and Laser -
Foundation for Research and Technology Hellas (FORTH),and Department
of Materials Science and Technology, University of Crete, Greece}

\author{Costas M. Soukoulis}
\affiliation{Ames Laboratory and Department of Physics and
Astronomy, Iowa State University, Ames, Iowa 50011, USA}
\affiliation{Institute of Electronic Structure and Laser -
Foundation for Research and Technology Hellas (FORTH),and Department
of Materials Science and Technology, University of Crete, Greece}

\begin{abstract}
We demonstrate experimentally and numerically that metamaterials
based on bilayer cross wires give giant optical activity,
circular dichroism, and negative refractive index. The presented chiral
design offers a much simpler geometry and more efficient way to
realize negative refractive index at any frequency. We also
developed a retrieval procedure for chiral materials which works
successfully for circularly polarized waves.
\end{abstract}
\maketitle

Recently, chiral metamaterials are proposed as an alternative to
realize negative refractive index. \cite{tretyakov_J_Elec_wave_2003,Pendry_Science_chiral_2004,tretyakov_FPNA_chiral_2005}
Chiral metamaterials are metamaterials made of unit cells without
symmetry planes. It has been shown that backward
waves exist in chiral media. \cite{tretyakov_J_Elec_wave_2003,
tretyakov_FPNA_chiral_2005} A chiral material slab can focus incident EM beams and can be used as a perfect lens. \cite{Jin_opex_chiral_2005,monzon_PRL_chiral_123904_2005} In 2004,
the canonical helix \cite{tretyakov_FPNA_chiral_2005} and the
twisted Swiss-role metal structures \cite{
Pendry_Science_chiral_2004} for microwave applications have been
discussed as possible candidates to achieve negative refractive
index. Later, the bilayer rosette-shaped chiral metamaterial is
proposed and fabricated at microwave frequency
\cite{rogacheva_PRL_177401_2006,PRB_Eric_chiral_2009} and at optical
regime.
\cite{plum_APL_223113_2007,vallius_APL_234_2003,Decker_OPL_chiral_2007}
This metamaterial exhibits a very strong rotary power in the microwave,
near-infrared and visible spectral ranges. In the microwave spectral
region, in terms of rotary power per wavelength, the bilayer
structure rotates five orders of magnitude stronger than a
gyrotropic crystal of quartz \cite{rogacheva_PRL_177401_2006}. It
has been shown that the strong gyrotropy originates from the
magnetic resonance of the bilayer metallic structure with
anti-parallel current flowing in the bilayer metal wires.
\cite{svirko_APL_chiral_498_2001} In this sense, the bilayer chiral
structure is the chiral version of the short wire pair
\cite{optleter_shalaev_2005, PRB_zhou_cwp,OL_Dolling_wire_pair} type
of metamaterials. The planar chiral structure
\cite{fedotov_PRL_chiral_planar_167401_2006,bai_PRA_023811_2007} and
chiral photonic crystal \cite{Thiel_OPL_chiral_PC_2007} are
proposed and fabricated. More recently, a chiral SRR consisting of
double layers of SRRs is proposed to provide negative refractive
index \cite{jelinek_PRB_chiral_205110_2008}.

In this letter we demonstrate, experimentally and numerically, the
negative refraction using the bilayer cross-wire design. We show
the negative refraction originates from the 3D chiral properties
of the bilayer cross wires. Unlike conventional negative index
material designs, such as the split-ring resonator type design
\cite{PRL_Smith_first_NIM} and fishnet designs,
\cite{science_soukoulis_2006} the chiral negative index material does
not require simultaneously negative permittivity and permeability,
and, therefore, the chiral design can offer much simpler geometry
and a more efficient way to realize negative refraction index. Due
to the asymmetric geometry of the cross-wire pairs, the cross
coupling between the magnetic field and electric field happens at
the chiral resonances and provides strong chirality around the resonance
frequencies. Further study shows chiral resonances are either electric
resonance or magnetic resonance of short wire pairs. The negative refractive
index of chiral metamaterial arises from this strong chirality, which splits the refractive
indices, $n_\pm$, of the two circularly polarized waves and makes the
refractive index of one circular polarization become negative.
\cite{Pendry_Science_chiral_2004} Our study shows that two resonance
modes exist for the cross-wire pairs. The resonance mode at the lower
frequency is a magnetic resonance mode with anti-parallel currents,
while the resonance mode at the higher frequency is an electric
resonance with parallel currents.
\begin{figure}[htb]\centering
  \includegraphics[width=8cm]{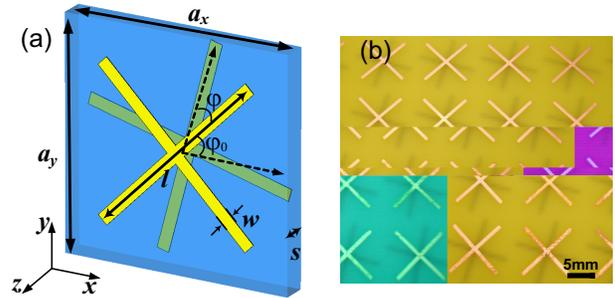}\\
  \caption{(Color online)
  (a) Schematic representation of one unit cell of the cross-wire structure.
  (b) Photograph of one side of a fabricated microwave-scale cross-wire sample.
  The geometry parameters are given by $a_x=a_y=15$mm, $l=14$mm, $w=0.8$mm, $s=1.6$
  mm, $\phi_0=45^\circ$, and $\phi=30^\circ$. The thickness of the copper is $t_m$=36$\mu$m.
  \label{fig_geom}}
\end{figure}

We develop a retrieval procedure adopting the uniaxial bi-anisotropic
model to calculate the effective parameters, $\mu$, $\epsilon$,
$\kappa$ and $n_\pm$, of the chiral metamaterial design. We prove
the existence of the negative index originating from the
chirality, $\kappa$ of the cross-wire metamaterial. As a comparison,
the non-chiral version of the cross-wires pair design does not show any
negative refractive index. Finally, we study the current
distribution for both the electric and the magnetic resonances, and
determine electric and magnetic resonances are the
symmetric and antisymmetric resonance modes of the coupled short
wire resonances, respectively.

%
The layout of the proposed structure is shown in Fig. 1.
 An 18 $\times$ 14 array of cross wires is patterned on a double side
copper-clad FR-4 board. The overall size of our sample is $279
\times 216$mm$^2$. The dielectric constant of the FR-4 board is
$\epsilon_r$=4.5+0.15i. The the properties of the cross-wires shown in Fig. 1 are characterized, using simulations and microwave measurements, and these results are used to determine the expected properties of metamaterials built from the cross-wire building blocks.  Simulations of cross-wire structures are achieved with CST Microwave Studio (Computer Simulation Technology GmbH, Darmstadt, Germany), which uses a finite element method to determine reflection and transmission properties.  In the simulations, the periodic boundary conditions are applied to a single unit cell shown in Fig. 1(a). Since the eigen solutions of the electromagnetic (EM) wave in chiral
materials are two circularly polarized EM waves, \cite{Elec_mag_wave_Kong,Monzon_IEEE_microwave_chiral_1990} i.e., the
right-handed circularly polarized wave (RCP,+) and the left-handed
circularly polarized wave (LCP,-), correspondingly, four
transmission coefficients, \cite{notation_T} $T_{++}$, $T_{-+}$, $T_{+-}$, and
$T_{--}$, are obtained to fully characterize the response of
the chiral metamaterials.  In the experiment, the transmission coefficient is measured by HP 8364B network analyzer with two Narda standard horn antennas.
Four linear transmission coefficients, $T_{xx}$, $T_{yx}$, $T_{xy}$, and
$T_{yy}$, are measured and the circular transmission coefficients,
$T_{++}$, $T_{-+}$, $T_{+-}$, and $T_{--}$ are converted from the
linear transmission coefficients using the following equation,
{\small
\begin{eqnarray}\label{equ_T_cir_lin}
&&\hspace{-0.2in}\left(%
\matrix{%
T_{++} & T_{+-}\cr%
T_{-+} & T_{--}}%
\right)=%
\frac{1}{2}\times\nonumber\\%
&&\hspace{-0.2in}\Biggl(%
\matrix{%
(T_{xx}+T_{yy})+\mathrm{i}(T_{xy}-T_{yx}) & (T_{xx}-T_{yy})-\mathrm{i}(T_{xy}+T_{yx})\cr%
(T_{xx}-T_{yy})+\mathrm{i}(T_{xy}+T_{yx}) & (T_{xx}+T_{yy})-\mathrm{i}(T_{xy}-T_{yx})}%
\Biggr)\hspace{0.2in}%
\end{eqnarray}
}

\begin{figure}[htb]\centering
  \includegraphics[width=8cm]{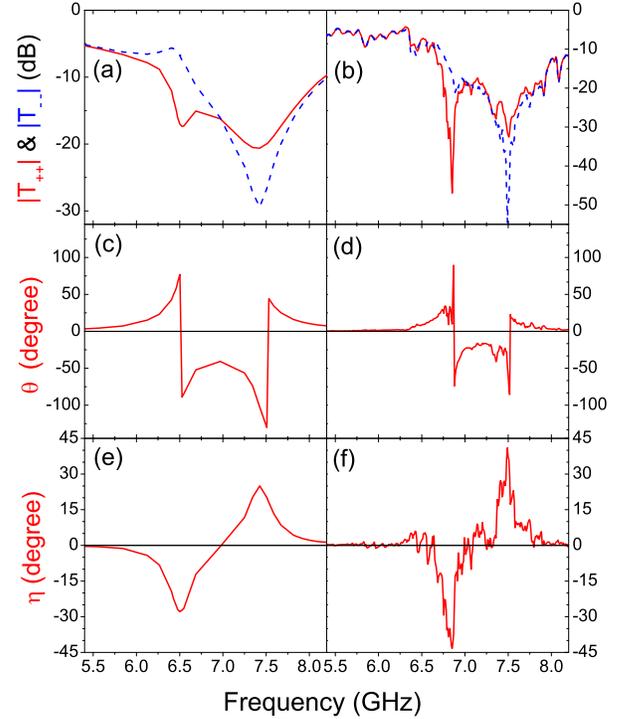}\\
  \caption{(Color online)
  (a) and (b): Simulated and measured magnitude of the transmission coefficients for the
  right circularly polarized, $|T_{++}|$ (red solid), and the left circularly polarized,  $|T_{--}|$ (blue dashed), electromagnetic
  wave, respectively. (c) and (d): Simulated and measured polarization azimuth rotation, $\theta$, respectively. (e) and (f): Simulated and measured
  ellipticity angle, $\eta$, respectively.\label{fig_T_theta}}
\end{figure}

Figures \ref{fig_T_theta}(a) and (b) show the simulated and measured
transmission coefficients, $T_{++}$ and $T_{--}$, as a function of
frequency, respectively (the cross coupling transmission, $T_{-+}$
and $T_{+-}$, are negligible, not shown here). Due to the
asymmetric geometry along the propagating direction, the
transmission responses for RCP and LCP split into two curves. Notice
two resonance dips are observed at frequencies, $f=6.5$ and 7.5
GHz, in both $T_{++}$ and $T_{--}$ curves. For the first resonance
at 6.5 GHz, the transmission dip for RCP, $T_{++}$, is much deeper
than that for LCP, $T_{--}$, which implies the resonance for RCP is
much stronger than LCP. While, for the second resonance at 7.5 GHz,
we observe the resonance for LCP is much stronger than RCP. Use the
standard definitions \cite{Elec_mag_Jackson} of the polarization
azimuth rotation,
$\theta=[\mathrm{arg}(T_{++})-\mathrm{arg}(T_{--})]/2$, and the
ellipticity,
$\eta=\frac{1}{2}\mathrm{arcsin}\left(\frac{|T_{++}|-|T_{--}|}{|T_{++}|+|T_{--}|}\right)$,
of elliptically polarized light, we calculate the polarization
changes of a linearly polarized wave incident on the cross wire
structures. The simulated and measured azimuth rotation, $\theta$,
and ellipticity, $\eta$, are presented in Fig.
\ref{fig_T_theta}(c),(e) and (d),(f), respectively. At the resonance
frequencies of 6.5 and 7.5 GHz, the azimuth rotation and
ellipticity reach their maximum values, ($\theta=-89^\circ$ ,
$\eta=-28^\circ$) and ($\theta=-130^\circ$ , $\eta=28^\circ$),
respectively. In the region between two resonance dips (around 6.9
GHz), the region with low loss and nearly zero
dichroism, we observes a polarization rotation of $-40^\circ$ with
$\eta\approx 0$, about four times larger than the value
reported using similar bi-layer chiral metamaterial designs, such as bilayer of twisted rosettes. \cite{rogacheva_PRL_177401_2006,PRB_Eric_chiral_2009} The sign change of $\eta$ at 7.0 GHz
reflects the different frequency dependence between the magnitude of the transmission $|T_{++}|$ and $|T_{--}|$.
As a consequence, the outgoing wave of a linear polarized incident wave below and above 7.0 GHz have different handedness.
Our numerical simulations show that the differences between $|T_{++}|$ and $|T_{--}|$ result from the different amounts of loss as the RCP and LCP waves pass through the cross wire structures.

To study the effective parameters of chiral metamaterials,
we develop a retrieval procedure based on a uniaxial bi-anisotropic
model.\cite{PRB_Eric_chiral_2009} The cross wire pairs design can be modeled as a reciprocal
uniaxial bi-anisotropic medium and the constitutive equation is
given by
%
{\small
\begin{equation}\label{equ_bi_iso}
\left(%
\matrix{%
\bf{D}\cr%
\bf{B} }%
\right)%
=
\left(%
\matrix{%
\epsilon_0\epsilon & -\mathrm{i}\kappa/c_0\cr%
\mathrm{i}\kappa/c_0 & \mu_0\mu}%
\right)%
\left(%
\matrix{%
\bf{E}\cr%
\bf{H}}%
\right),%
\end{equation}
}
where $\epsilon_0$, $\mu_0$, and $c_0$ are the permittivity,
permeability, and the speed of light in a vacuum, respectively.
Assuming $\exp(-\mathrm{i}\omega t)$ time dependent, the eigen
solution of the electromagnetic wave in bi-isotropic media is circular
polarized plane waves, and the polarization is either left-handed
circular polarized (LCP) or right-handed polarized (RCP). The
refractive indices for LCP and RCP are given by
\cite{Elec_mag_wave_Kong,Monzon_IEEE_microwave_chiral_1990}
%
\begin{equation}\label{equ_chiral_n}
n_\pm=\sqrt{\epsilon\mu}\pm\kappa,
\end{equation}
where $(+)$ and $(-)$ denote RCP and LCP, respectively. Both LCP and RCP have the
same impedance given by $z/z_0=\sqrt{\mu/\epsilon}$, where $z_0$ is
the impendence of the vacuum.

From Eq. (\ref{equ_chiral_n}), one can immediately see that $n<0$
for one of the polarizations, if $\kappa$ is large enough, such that
$\sqrt{\epsilon\mu}<\kappa$. It has been shown that the polarization
azimuth rotation, $\theta$, is proportional to the chiral parameter, $\kappa$. Specifically, $\theta=\kappa k_0
d$, where $k_0$ and $d$ are the wave vectors in a vacuum and the
thickness of the bi-isotropic slab. Thus, the chiral materials with
large rotary power, such as the bilayer structures and chiral SRRs,
possibly possess the negative refractive index. However, it is not
trivial to obtain negative refractive index by a chiral design with
a large azimuth rotation. One should keep in mind that the large value
of azimuth angle, $\theta$, happens at the resonances, where
$\sqrt{\epsilon\mu}$ is also large, so $\kappa$ should be large
enough to overcome the large magnitude of $\sqrt{\epsilon\mu}$ to
achieve negative $n$.

\begin{figure}[htb]\centering
  \includegraphics[width=7cm]{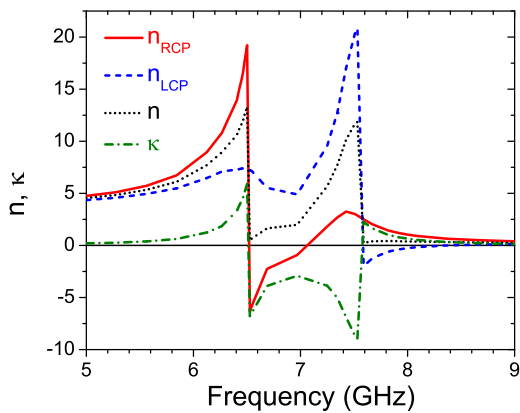}\\
  \caption{(Color online)
  The retrieved effective refractive index for the right circularly polarized EM wave, $n_+$ (red solid),
  and the left circularly polarized EM wave, $n_-$ (blue dashed). The black dotted curve shows the
  refractive index calculated by the permittivity and the
  permeability, $n=\sqrt{\epsilon\mu}$, and the green dash-dotted
  curve shows the chiral parameter, $\kappa$.\label{fig_n}}
\end{figure}
In Fig. \ref{fig_n}, we present the refractive index for RCP and
LCP, $n_+$, $n_-$, the conventional definition of refraction index,
$n=\sqrt{\epsilon\mu}$, and the chiral parameter, $\kappa$, which are extracted
from simulation results using the retrieval procedure.
A thickness, $d=s+2t_m=1.672$mm, is used when
we extract these effective parameters.
Notice
that $n$ (black dotted curve) is positive through the entire
frequency range from 5 to 9 GHz. However, $n_+$ (red solid) is
negative from 6.5 to 7.0 GHz and $n_-$ (blue dashed) has a negative
region from 7.6 to 8.2 GHz. The chiral parameter (green dash-dotted
curve) shows two resonances at 6.5 and 7.5 GHz, respectively. Above
the first resonance frequency, $\kappa$ is negative between 6.5 to 7.0
GHz, which leads to $n_+=\sqrt{\epsilon\mu}+\kappa<0$ between
6.5 to 7.0 GHz . Similarly, above the second resonance frequency,
$\kappa$ is positive and results in
$n_-=\sqrt{\epsilon\mu}-\kappa<0$ between 7.6 to 8.2 GHz. It is
clear the negative refractive index for RCP and LCP originates
from the chiral parameter, $\kappa$. The observed negative
refractive index $n_+=-2.5$ at 6.8 GHz has the figure of merit,
FOM=$|\mathrm{Re}(n)/\mathrm{Im}(n)|$=0.75, and $n_-=-1$ at 7.8
GHz has FOM=0.5. The figure of merit is relatively low compared to the
conventional negative index material designs such as SRRs or fishnet.
The low FOM is due to the high loss of this chiral metamaterial design.
Further study shows the loss mainly originates from the lossy dielectric
spacer. If low loss dielectric materials is used in the cross wire design,
the FOM can improve substantially. For instance, in our numerical simulations,
 we obtained FOM$\approx$10 using dielectric spacer with $\epsilon_r$=4.5+0.005i.

We also calculate the $n_+$, $n_-$, $n$ and $\kappa$ for a
non-chiral cross-wire pairs with the mutual twisted angle $\phi=0$.
The results show $n_+=n_-=n>0$, and the chiral parameters, $\kappa=0$ through the entire frequency range from 5 to 9.0 GHz. Thus, we confirme the negative refractive index
observed in Fig. \ref{fig_n} is due to the chirality introduced by
the asymmetry of the cross-wire pairs.

\begin{figure}[htb]\centering
  \includegraphics[width=7cm]{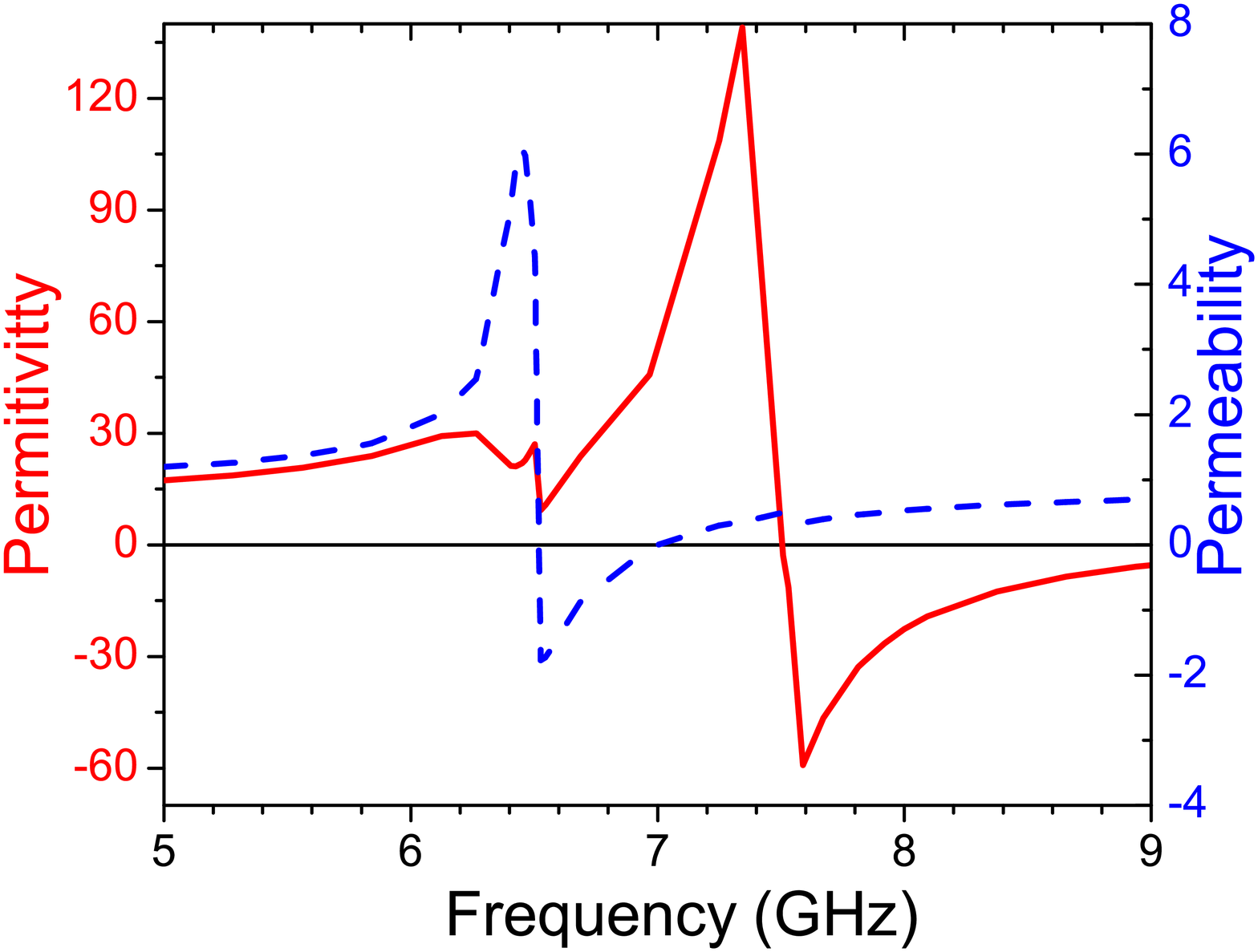}\\
  \caption{(Color online)
  The real part of the effective permittivity (red solid) and the effective permeability (blue dashed), extracted from
  the simulation data.
  \label{fig_eps_mu}}
\end{figure}
In Fig. \ref{fig_eps_mu}, we present the real parts of the
permittivity, $\epsilon$, and the permeability, $\mu$, as a function
of frequency for the asymmetric cross wire pairs. Two resonances are
observed in $\mu$ (blue dashed) at 6.5 and $\epsilon$ (red
solid) at 7.5 GHz, respectively. This demonstrates the resonance
observed in Figs. \ref{fig_T_theta} and \ref{fig_n} at 6.5 GHz is a
magnetic resonance and the resonance at 7.5 GHz is an electric
resonance. The magnetic resonance gives the negative $\mu$ between
6.5 to 6.9 GHz, and the electric resonance give a negative
$\epsilon$ between 7.6 to 9 GHz. There is no overlap region of
negative $\epsilon$ and $\mu$. Therefore,
$\mathrm{Re}(n)=\mathrm{Re}(\sqrt{\epsilon\mu})>0$, which is
consistent with the observation of $\mathrm{Re}(n)>0$ shown in Fig. \ref{fig_n}.
\begin{figure}[htb]\centering
  \includegraphics[width=8cm]{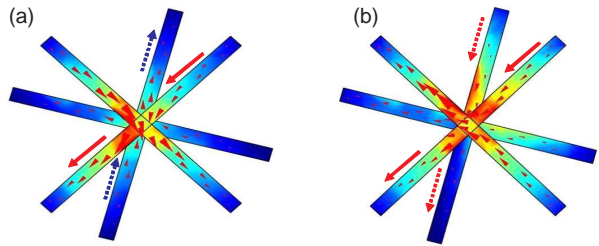}\\
  \caption{(Color online)
  The simulated current density distribution for the right circularly polarized EM wave at 6.5 GHz
  (a) and for the left circularly polarized EM wave at 7.5 GHz (b). The cones (in red) show the direction and
  magnitude of the current density. The magnitude of the current density
  is also shown by the color on the wires, with red and blue corresponding to largest and smallest values.
  \label{fig_current}}
\end{figure}

In order to understand the mechanism of the resonances for the
cross wire pair design, we studied the current density distribution
as shown in Fig. \ref{fig_current}. Notice, at the magnetic
resonance, the anti-parallel current exists on the top and bottom
layer of cross-wire pairs (Fig. \ref{fig_current}(a)), which is an
asymmetric resonance modes. In Fig. \ref{fig_current}(b), one can
see parallel currents flowing on the two layers of wires, which is a
symmetric resonance mode.  The current distribution shows that the
cross wire pairs can be viewed as a chiral version of the short wire
pairs, \cite{optleter_shalaev_2005,
PRB_zhou_cwp,OL_Dolling_wire_pair} which has similar current
distributions in the symmetric and asymmetric resonance modes.

In summary, we experimentally demonstrated bilayer cross wires
with very adaptable properties, including negative index of
refraction due to chirality, giant optical activity and very large
circular dichroism. We performed numerical simulations that give
evidence that the negative refractive index is due to the chiral
nature of the metamaterial and not from the simultaneous negative $\epsilon$
and $\mu$ as for conventional negative index materials.
In addition, we developed a retrieval procedure that works well for chiral
metamaterials. The geometry of the
cross wire design is simple and easy-to-fabricate, and therefore,
is more suitable for optical frequency applications
compared to other types of bi-layer chiral metamaterial designs.

Work at Ames Laboratory was supported by the Department of Energy
(Basic Energy Sciences) under contract No. DE-AC02-07CH11358. This
work was partially supported by the Department of Navy, Office of
the Naval Research (Award No. N00014-07-1-0359), European Community
FET project PHOME (Contract No. 213390) and AFOSR under MURI grant
(FA 9550-06-1-0337). The author Jianfeng Dong gratefully acknowledges
 support of the W.C. Wong Education Foundation, Hong Kong,
 the National Basic Research Program (973) of China
 (Grant No. 2004CB719805) and the National Natural Science Foundation
 of China (Grant No.60777037). \\

\begin{thebibliography}{24}
\bibitem[{not()}]{notation_T}
\bibinfo{journal}{In these notations, the first and second subscripts indicate
  the output and input signal polarization, respectively. For example, $T_{+-} =
  E^t_+ /E^i_-$ , where $E^i_-$ is the input LCP electric field and $E^t_+$ is
  the transmitted RCP electric field}.
\end{thebibliography}


\end{document}